\documentstyle{ioplppt}

\def\rmi{{\rm i}}
\def\rme{{\rm e}}
\def\rmd{{\rm d}}

\def\Re{\hbox{\rm Re}}
\def\Im{\hbox{\rm Im}}

\def\calO{{\cal O}}

\begin{document}


\title{Some properties of 
eigenvalues and 
eigenfunctions of the
cubic oscillator with imaginary coupling constant}[Cubic oscillator with imaginary coupling constant]

\author{G. Andrei Mezincescu\dag
\ftnote{2}{E-mail: mezin@alpha1.infim.ro}}

\address{\dag INFM, 
C.P. MG-7, R-76900 Bucure\c sti -- M\u agurele, Rom\^ania, 
and\\ Centrul de Cercet\u ari Avansate de Fizic\u a al  Academiei
Rom\^ane, Bucure\c sti, Rom\^ania}

\begin{abstract}
Comparison between the exact value of the spectral zeta function, 
$Z_{H}(1)=5^{-6/5}\left[3-2\cos(\pi/5)\right]\Gamma^2(1/5)/\Gamma(3/5)$,
and the results of numeric and WKB calculations supports the conjecture by 
Bessis that all the eigenvalues of this PT-invariant hamiltonian are real.
For one-dimensional Schr\"odinger operators with complex 
potentials having a monotonic imaginary part, the eigenfunctions (and the 
imaginary parts of their logarithmic derivatives) have no real zeros.
\end{abstract}


\section{Introduction}

Some years ago Bessis (1995) had conjectured that all the eigenvalues of the 
cubic oscillator with purely imaginary coupling constant,
\begin{equation}
H=-\frac{\rmd^2}{\rmd x^2} + \rmi\,g\,  x^3; \qquad \Im(g)=0,
\label{1.1}\end{equation}
were real. Numerical and WKB investigations have found only real eigenvalues
(Bessis, 1995, Bender and Boettcher 1998ab, Bender \etal 1999), 
although a rigorous proof is still lacking. 

The operator \eref{1.1}, defined on the intersection of the domains
of ${\rmd^2}/{\rmd x^2}$ and $x^3$ is closed and has a compact,
Hilbert-Schmidt resolvent (Caliceti \etal 1980, Caliceti 1999).
Thus, its spectrum consists only of eigenvalues, which can acumulate
only near infinity.  Since its numerical range lies in the right complex
half-plane [$\Re(z)>0$], the real parts of the eigenvalues are positive
(Kato 1966).

A Schr\"odinger operator 
\begin{equation}
H=-\frac{\rmd^2}{\rmd x^2} +U(x),
\label{1.2}\end{equation}
is called PT-invariant if the potential satisfies
\begin{equation}
U(x)=\overline{U(-x)}.
\label{1.2a}\end{equation}
Here, and in the following, we use the overline to denote complex conjugation.
Although \eref{1.1} is not Hermitean (symmetric), it is PT-invariant, but not
separately invariant under the spatial inversion P,
[$\phi(x)\rightarrow\phi(-x)$], or the time reversal T,
[$\phi(x)\rightarrow\overline{\phi(x)}$], which map it on its adjoint $H^*$.

Adding harmonic term to the potential in \eref{1.1}, we obtain 
\begin{equation}
U(x)=\rmi g x^3+\mu x^2;  \qquad  \mu\ge 0.
\label{cubg}\end{equation}
The PT-invariant operator defined by \eref{1.2} and \eref{cubg} has been
intensely investigated. Caliceti \etal (1980) have shown that each eigenvalue
of the harmonic oscillator (to which it reduces for $g=0$) stays real for
sufficiently small real $g$ or, equivalently, for sufficiently large $\mu$
(Delabaere and Pham 1999).

The resolvent $(z-H)^{-1}$ of the one-dimensional Schr\"dinger operator
\eref{1.2} is an integral operator for all $z$ in the resolvent set.
Its kernel, $G_z(x,y)$ can be expressed through 2 solutions that satisfy
the boundary condition at {\it only one} end of the interval
(Naimark 1964, Caliceti \etal 1980).
Let $f_+(x;z)$ [ $f_-(x;z)$ ] be the (unique, up
to a multiplicative constant) solution of the equation
\begin{equation}
(z-H)f=f^{\prime\prime}+ [z-U(x)]f=0, 
\label{1.3}\end{equation}
which, together with its first derivative, is locally L$^2$ near $+\infty$
(respectively $-\infty$).
One can readily check that the kernel is
\begin{equation}
G_z(x,y)= 
\frac{\theta(x-y)f_+(x;z)f_-(y,z)+\theta(y-x)f_+(y;z)f_-(x;z)}
{W[f_+,f_-](z)}.
\label{1.4}\end{equation}
Here $\theta(x)$ is the Heaviside function (characteristic function of 
the positive half-axis), and
\begin{equation}
W[f_+,f_-](z)=f_+^{\prime}(x;z)f_-(x;z)-f_-^{\prime}(x;z)f_+(x;z),
\label{1.5}\end{equation}
is the ($x$-independent) Wronskian of $f_+$ and $f_-$.
 
Generally speaking, a PT-invariant Schr\"odinger operator \eref{1.2},
\eref{1.2a} can have any kind of spectrum, as long as it is invariant
with respect to complex conjugation.
For example, let 
\begin{equation}
U_{g,a}= \rmi g \tanh(x/a), \hbox{~with~} g,\, a >0.
\label{tanh}\end{equation}
This PT-invariant potential of can be thought as an interpolation
between the potential $\rmi Fx$ of an imaginary electric field
($g=aF,~a\to+\infty$) and the imaginary step potential
$\rmi g\,\hbox{sign}(x)$ (in the limit $a\to 0$). 
Define $H_{g,a}$ by \eref{1.2} with the potential \eref{tanh} 
on the domain of the second derivative. The solutions of the eigenvalue 
equation for \eref{tanh} can be obtained by analytic continuation
from those given in (Landau and Lifschitz 1977).

The operator  $H_{g,a}$ has no eigenvalues. Its spectrum, which does not 
depend on $a$, is the union of two half-lines parallel to the real axis: 
$\left\{\pm\rmi g +t; ~t\ge 0\right\}$.
In the imaginary electric field limit, the lines move to infinity, so that 
the spectrum of the limiting operator reduces to the point at infinity 
(see e.g. Herbst 1979).

Another instructive exactly soluble model is the 
$PT$-invariant "complexified harmonic oscillator" with the potential
\begin{equation}
U_\lambda(x)=\left\{
  \begin{array}{ll} 
    \rme^{\rmi\lambda} x^2,      &\mbox{if $x>0$ };\\
    \rme^{-\rmi\lambda} x^2,      &\mbox{if $x<0$ };
  \end{array}
\right. 
\label{sq}\end{equation}
where $|\lambda|\le\pi/2$. For $\lambda=0$, (\ref{sq})
coincides the potential of the harmonic oscillator, while for
$\lambda=\pm\pi/2$ it is purely imaginary and antisymmetric, mimicking the
cubic potential in (\ref{1.1}). Let us also note that, in the limit $\lambda\to 0$
the perturbation \eref{sq} on the harmonic oscillator is somewhat milder than
for the $\epsilon\to 0$ limit of the potential $-(\rmi x)^{2+\epsilon}\approx 
x^2[1+\epsilon\ln(|x|)+\rmi\frac{\epsilon\pi}2{\rm sign}(x)$, considered by 
Bender and Boettcher(1998a). 

The operator $H_\lambda$, defined on $D(\rmd^2/\rmd x^2)\cap D(x^2)$,  has
compact (Hilbert-Schmidt) resolvent.  But, for any $\lambda\ne 0$ only a 
{\it finite} number of the eigenvalues are {\it real}.  The rest come in 
complex-conjugated pairs.

In the next section, we will compute exactly one value of the spectral zeta function
($Z_ H(1)$), the sum of the inverse eigenvalues, and will compare the
result with numerical and WKB calculations for the real eigenvalues of \eref{1.1},
obtaining a remarkable coincidence.
Since the real parts of any presumed complex eigenvalues must be positive,
this places stringent bounds on the values of the maximum value of the ratio 
$|\Re(E_\alpha)/E_\alpha^2|$ for complex eigenvalues.
In section 3 we will  also show that the eigenfunctions (and their first
derivatives) have no zeros on the real axis. This property holds for
all potentials with {\it monotonic} imaginary part.

\section{Sum of inverse eigenvalues}

In our case we can set $g=1$ without altering generality. The eigenvalue
equation \eref{1.3} reads
\begin{equation}
f^{\prime\prime}+\left(z-\rmi x^3\right)f=0.
\label{2.1}\end{equation}
Generally speaking, the solutions of (\ref{2.1}) cannot be expressed
in terms of the classical special functions.
This is possible only for $z=0$ (Gradshtein and Ryzhik 1965, Caliceti \etal
1980).  Then, we can readily verify that
\begin{equation}
\fl f_+(x;0)=\left\{
\begin{array}{l}
x^\frac 12 K_\frac 15 \big(\frac 25 \rme^\frac{\rmi\pi}{4} x^\frac 52\big)
,\qquad x>0;\\
\frac{\pi |x|^\frac 12}{2\sin\frac{\pi}5} \Big[
\rme^{-\frac{\rmi\pi}{10}}
I_{-\frac 15}\big(\frac 25 \rme^{-\frac{\rmi\pi}{4}} |x|^\frac 52\big)
+\rme^{\frac{\rmi\pi}{10}}
I_{\frac 15}\big(\frac 25 \rme^{-\frac{\rmi\pi}{4}} |x|^\frac 52\big)
\Big],\qquad x<0;
\end{array}
\right.
\label{2.2}\end{equation}
and
\begin{equation}
f_-(x;0)=\overline{f_+(-x;0)},
\label{2.3a}\end{equation}
are a pair of PT-conjugated solutions of the \Eref{2.1}, and
that $f_+(x;0)$ [$f_-(x;0)$] and their first derivatives are locally
L$^2$ near  $+\infty$ [$-\infty$].
Here, $I_\nu$ and $K_\nu$ denote the Bessel function of the first kind
and imaginary argument, respectively Macdonald's function.

Although this is not immediately apparent from the representation \eref{2.2},
the solutions $f_+(x;0)$ and $f_-(x;0)$ are integer functions of $x$.
The first terms of the McLaurin series for $f_+(x;0)$ are
\begin{equation}
f_+(x;0)=
\frac{\pi }{2\sin\frac{\pi}{5}} \bigg[
\frac{5^{-\frac 15}\rme^{-\frac{\rmi\pi}{20}}}{\Gamma\left(\frac 45\right)} -
\frac{5^{\frac 15}\rme^{\frac{\rmi\pi}{20}}}{\Gamma\left(\frac 65\right)} x
+\calO(x^5)\bigg].
\label{2.4}\end{equation}
From (\ref{2.4}) and \eref{2.3a} we get the $z=0$ value of the Wronskian
(\ref{1.5})
\begin{equation}
W[f_+,f_-](0)=-\frac{5\pi}{4\sin\frac{\pi}{10}}.
\label{2.5}\end{equation}

The asymptotic behaviour of $f_+(x;0)$ near $\pm\infty$ is given by
\begin{equation}
\fl
f_+(x;0)\approx\left\{
\begin{array}{ll}
\frac{\sqrt{5\pi}}{2x^\frac 34}
\exp\big(-\frac 25 \rme^\frac{\rmi\pi}{4} x^\frac 52
-\frac{\rmi\pi}{8}\big)\left[1+\calO(x^{-\frac 52})\right], & x\to+\infty;\\
\frac{\sqrt{5\pi}}{|x|^\frac 34\sin\frac\pi{10}}
\exp\big(\frac 25 \rme^{-\frac{\rmi\pi}{4}} |x|^\frac 52
+\frac{\rmi\pi}{8}\big)\left[1+\calO(|x|^{-\frac 52})\right], & x\to-\infty.
\end{array}
\right.
\label{2.6}\end{equation}

Using the asymptotic behaviour \eref{2.6}, Caliceti \etal (1980) have shown
that the kernel $G_0(x,y)$  belongs to the Hilbert-Schmidt class.
\begin{eqnarray}
||G_0||_{\rm HS}=
\int_{-\infty}^{+\infty}\int_{-\infty}^{+\infty}\rmd x \rmd y
\vert G_0(x,y)\vert^2 < \infty.  
\label{2.7} \end{eqnarray}
Here we will note that the diagonal elements of the continuous 
in $(x,y)$ Hilbert-Schmidt kernel $G_0(x,y)$ is absolutely summable, 
\begin{equation}
\int_{-\infty}^{+\infty}\rmd x 
\vert G_0(x,x)\vert<\infty,
\label{2.10}\end{equation}
since the continuous integrand is summable near both infinities,
\begin{equation}
\vert f_+(x;0)f_-(x;0)\vert\approx \calO\left(|x|^{-3/2}\right),
\hbox{\rm when} x\to\pm\infty.
\label{2.10a}\end{equation}

The trace is equal to the sum of all the inverse eigenvalues
\begin{equation}
Z_H(1)=\sum_j\frac 1{E_j}=-\int_{-\infty}^{+\infty}\rmd x G_0(x,x),
\label{2.11}\end{equation}
which is finite. 
Analogous formulae can be written for the zeta functions of integer
arguments $s>1$.
Taking into account the PT-conjugacy of $f_+$ and $f_-$, we obtain
\begin{eqnarray}
\fl Z_ H(1)=\frac{2\sin\frac \pi{10}}{5\pi}
\Re\left[\int_{0}^{+\infty}\rmd x f_+(x;0)f_-(x;0)\right] = \nonumber\\
\left(\frac 25\right)^\frac 65 \int_{0}^{+\infty}\rmd t ~t^{-\frac 15}
K_{\frac 15}(t) \left[I_{-\frac 15}(t)+
\frac{\cos(3\pi/10)}{\cos(\pi/10)}
I_{\frac 15}(t)\right]=\nonumber\\
\frac{\Gamma^2\left(\frac 15 \right)}{5^{6/5}\Gamma\left(\frac 35\right)}
\left[3-2\cos\frac{\pi}5\right]\approx 2.835094933. 
\label{2.12} \end{eqnarray}
In the last integral we rotated the contour of integration by $-\pi/10$, 
making  the substitution
$x=\left(5t/2\right)^{\frac 25}\rme^{-\rmi\pi/10}$,  which leads to
tabulated integrals (Gradshtein and Ryzhik 1965, {\bf 6.576}.5).

Bender and Boettcher (1998a) gave a WKB estimate for the large
eigenvalues of \eref{2.1}
\begin{equation}
E_j\approx E_j^{(qc)}=
\left[5\sqrt{\frac{\pi}3}
\frac{\Gamma\big(\frac 56\big)}{\Gamma\big(\frac 13\big)}
\Big(j+\frac 12\Big)\right]^{6/5}.
\label{wkb}\end{equation}
Comparison of \eref{wkb} with the numerical results of Bender and Boettcher 
(1998a) shows that even for $j=0$ the error is about $5\%$.

The sum of all inverse WKB eigenvalues \eref{wkb} can be 
expressed in terms of the Riemann zeta function:
\begin{equation}
\fl Z_H^{(qc)}(1)=\sum_j\frac 1{E_j^{(qc)}}=
\left[5\sqrt{\frac{\pi}3}
\frac{\Gamma\big(\frac 56\big)}{\Gamma\big(\frac 13\big)}
\right]^{6/5}
\left(2^{\frac 65}-1\right)\zeta\left(\frac 65\right)\approx 2.885673793.
\label{qc}\end{equation}

We can correct the WKB approximation \eref{qc} for $Z_H(1)$ by taking the numerical
values of the first five eigenvalues from Table 1 of Bender and
Boettcher (1998a), who give their result to five digit precision. 
This yields $2.8351$, to be compared with $2.835094933$ from \eref{2.12}.
A remarkable coincidence!
The two values coincide within the five-digit acuracy of the
numerical eigenvalues.
Since the real parts of any presumed complex eigenvalues must be positive, 
no compensations can occur in the sum. If some complex conjugate eigenvalues 
$E_\alpha$ are present, then the largest  ratio
\begin{equation}
\max_\alpha \left|\frac{\Re(E_\alpha)}{E_\alpha^2}\right| <10^{-5}. 
\end{equation}
We consider this as strongly supportive for the conjectured (Bessis 1995) 
absence of complex eigenvalues.

\section{Nodeless character of the eigenfunctions}

As we can see from \eref{wkb}, the sum \eref{2.11} converges slowly. 
An improvement in the precision of the comparison between \eref{2.12} and
numerical result swill require precision computations of many eigenvalues. 
Bender and Boettcher (1998a) integrated the differential equation along 
the anti-Stokes lines in the lower complex half-plane, starting with large values of 
$|x|$, where the asymptotic behavior is attained, and matching the solutions 
at $x=0$. The following proposition on the properties of
the eigenfunctions could prove useful for simplifying numerical work.

Let $\phi_E$ be an eigenfunction corresponding to the eigenvalue $z=E$ in
(\ref{2.1}). Then, proceeding as in the case of proving current conservation
for the solutions of the stationary Schr\"odinger equation, we obtain
\begin{equation}
j_E^\prime(x)=\Im\left[\phi_E^\prime(x)\overline{\phi_E(x)}\right]^\prime
=[x^3 -\Im(E)]|\phi_E(x)|^2.
\label{2.13}\end{equation}
The current density associated to an eigenfunction is a decreasing
(increasing) function of $x$ if $x^3-\Im(E)<(>)0$. Since the limit 
$j_E(\pm\infty)=0$, the current density is strictly negative on $(-\infty,+\infty)$ 
and has a unique minimum at $x=\sqrt[3]{\Im(E)}$. Thus,
\begin{enumerate}
\item[a.] {\it The eigenfunctions of (\ref{2.1})
have no zeros on the real axis;}

\item[b.] {\it The imaginary part of their logarithmic derivatives,
$\Im\left[\phi_E^\prime(x)/\phi_E(x)\right]$, have constant
sign.}
\end{enumerate}
This proposition can be readily generalized to the eigenfunctions 
of \Eref{1.2} with a complex potential $U(x)$ having a {\it monotonic} 
imaginary part. 

Using the above proposition we can look for the eigenvalues of 
\eref{2.1} by a shooting method, integrating the Ricatti equation
satisfied by the logarithmic derivative $s(x)=f^\prime(x)/f(x)$
\begin{equation}
s^\prime+s^2+E-\rmi x^3=0.
\label{cu.8}\end{equation}
In the selfadjoint case, the eigenfunctions have nodes on the real axis,
which leads to singularities in the logarithmic derivative. Thus,
although the order of the equation can be lowered, this is not quite
helpful for numerical calculations.
In our case, the eigenfunctions corresponding to real eigenvalues
are PT-symmetric (eventually up to an irrelevant phase factor). 
This implies that if $s(x;E)$ is the logarithmic derivative of an 
eigenfunction, then $s(0;E)$ is purely imaginary.
The large $x$ asymptotic behaviour of the solution of \eref{cu.8} 
whose real part goes to $-\infty$ as $x\to +\infty$ is
\begin{equation}
s(x)\approx -   
\frac{1+\rmi}{\sqrt{2}}   
x^{\frac 32} -\frac 3{4x} 
+\frac{1-\rmi}{2\sqrt{2}}   
E x^{-\frac 32}
+\calO\left(x^{-\frac 72}\right); \qquad x\to+\infty.
\label{cu.9}\end{equation}
Now, choosing a sufficiently large $X$, where $s(X)$ for
the solution which decays at infinity is close to the asymptotic, we can 
integrate \eref{cu.8} from $X$ backwards to $x=0$. 
If $E$ is an eigenvalue, then 
\begin{equation}
\Re[s(0;E)]=0.
\label{cu.13}\end{equation}
High precision solutions of \eref{cu.13} could be obtained by a simple
Newton routine, starting at the WKB prediction \eref{wkb}.

The advantage of this approach is that we integrate a complex {\it first
order} equation instead of the original second order one. The requirement
that the imaginary part of $s$ is negative allows stopping the integration
if we detect a change of sign in $\Im(s,E)$. That energy cannot be an eigenvalue.
If $X$ is sufficiently large, the eigenvalues are relatively insensitive 
to small perturbations of the boundary condition at $X$. Practical
coonsiderations require us to start with the lowest value of $X$,
where the decaying solution is sufficiently close to the asymptotic behaviour 
\eref{cu.9}. An alternative approach for the large eigenvalues,  would be to 
obtain corrections to the leading order WKB result \eref{wkb}.

\ack  Partial support by the Academy of Finland under Project No 163394
is gratefully acknowledged. I thank Masud Chaichian for hospitality at
the Department of High Energy Physics, Helsinki University, where
part of this work has been done, and Daniel Bessis for discussions.

\References

\begin{harvard}

\item[] Akhiezer N I and Glazman I M 1966 {\it Theory of Linear Operators
in Hilbert Space} 2nd edition (Moscow: Nauka) (in Russian);
English translation 1993 (New-York: Dover)

\item[] Bender C M and Boettcher S 1998a \PRL {\bf 80} 5243

\item[]  \dash 1998b  \JPA {\bf 31} L273

\item[] Bender C M, Boettcher S and Meisinger P N 1999 \JMP
{\bf 40} 2201

\item[] Bessis D 1995 Private communication.

\item[] Caliceti E, Graffi S and Maioli M 1980 {\it Commun. Math. Phys.} 
{\bf 75} 51

\item[] Caliceti E 1999 {\it Distributional Borel summability of odd
anharmonic oscillators} Preprint mp\_arc 99-384; http://www.math.utexas.edu 

\item[] Delabaere E and Pham F 1998 \PL {\bf A250} 25

\item[] Gradshtein I S and  Ryzhik I M 1965 {\it Tables of Integrals,
Series and Products} 4th edition (New-York: Academic Press)

\item[] Herbst I 1979 {\it Commun. Math. Phys.} {\bf 64} 279

\item[] Kato T 1966 {\it Perturbation Theory for Linear Operators}
(Berlin: Springer)

\item[] Landau L D  and Lifschitz E M 1977 {\it Quantum Mechanics}
3rd edition (Oxford: Pergamon)

\item[] Naimark M A 1964 {\it Linear Differential Operators} Part II
(London: Harrap)

\item[] Reed M and Simon B 1972 {\it Methods of Modern Mathematical Physics}
Vol. 1 (New York: Academic Press)

\end{harvard}
\endrefs

\end{document}